\documentstyle[11pt,aaspp4]{article}
\begin{document}

\title{Variable stars in the field of the open cluster NGC 7789
\footnote{Based on observations obtained at Kitt Peak National 
Observatory, a division of NOAO which is operated by the Association of
Universities for Research in Astronomy, Inc. under cooperative agreement
with the National Science Foundation.}}

\author{B. J. Mochejska$^{1}$ and J. Kaluzny$^{1,2}$}
\affil{$^{1}$Warsaw University Observatory, Al. Ujazdowskie 4,
PL-00-478 Warszawa, Poland} 
\affil{\tt e-mail: mochejsk@sirius.astrouw.edu.pl, jka@sirius.astrouw.edu.pl} 
\affil{$^{2}$Copernicus Astronomical Center, 00-716 Warszawa, Bartycka 18}

\begin{abstract}
We present the results of our search for variable stars in the
intermediate-age open cluster NGC 7789. We have found 45 variable stars: 35
eclipsing binaries, five pulsating variables and five miscellaneous
variables. Most of the eclipsing binaries show W UMa type of variability,
with periods shorter than one day. Four systems exhibit unusual behavior:
two, V4276 and V6698, are probably RS CVn stars, another, V3283, is a
possible cataclysmic binary. The nature of the fourth binary, V2130, is
unclear: the system exhibits assymetric maxima. Among the pulsating
variables two, V3407C and V4805 are background RR Lyrae stars and one,
V6736, is a $\delta$ Scuti variable which is a blue straggler belonging to
the cluster. Some of the miscellaneous variables may have periods longer
than the five day timespan of our observations. We also present a
color-magnitude diagram for the NGC 7789 open cluster, fairly complete down
to $V\sim20$.  The relatively large number of variables found in the
comparison field (14 compared to 31 in the cluster field) implies that
objects not associated physically with the cluster can account for a
significant number of variables identified in the cluster field, as well as
in other globular and open clusters observed on a dense
background/foreground of disk stars.  

\end{abstract}

\section{Introduction}

We present the results of our search for variable stars in the rich
intermediate-age open galactic cluster NGC 7789, located at $l = 
115.\!\!\arcdeg49, b = -5.\!\!\arcdeg\;36$. The cluster has been the
subject of only one previous search for variable stars, conducted by Jahn
et al. (1995), which resulted in the discovery of 15 variables in the
central part of the cluster. 

The first extensive photometry of NGC 7789 was presented by Burbridge \&
Sandage (1958), providing magnitude and color determinations for nearly 700
stars within some $450\arcsec$ of the cluster center. Recent estimates
place the age of the cluster at about 1.6 Gyr (Gim et al. 1998). A
determination of the distance modulus by Jahn et al. (1995) based on the
lower limit of the red-giant clump on the color-magnitude diagram yields
$(m-M)_V = 12.3$.

In section 2 we provide general information on the location of the
monitored fields and the observations. Section 3 describes the reduction 
procedure applied to the collected data. In section 4 the method used for
selecting candidate variables is discussed briefly. The color-magnitude 
diagrams are shown in Section 5. We present the catalog of variables in 
section 6. In section 7 we apply an absolute brightness calibration for 
contact binary systems to try to assess their cluster membership. A brief 
discussion of the results is presented in section 8.

\section{Observations}

The data for this project was obtained with the Kitt Peak National
Observatory 0.9m telescope equipped with a Tektronix $2048\times2048$ CCD
(T2KA camera) on five consecutive nights, from October 17 to 21, 1992. The 
field of view was about $23\times23$ arcmin$^2$ with the scale of
$0.68\arcsec/pixel$.

Two fields were monitored, one centered on the NGC7789 open
galactic cluster($\alpha=23.\!\!^h952$, $\delta=56.\!\!\arcdeg71$).
The other field, intended for comparison, was offset, for the most part, 
in right ascension and had a small overlap with the cluster field
($\alpha=23.\!\!^h988$, $\delta=56.\!\!\arcdeg64$). 

Useful data was collected during five to six hours each night, except for
the night of October 19, when only a few good frames were obtained, giving
a total of 22.5 hours of monitoring. For the cluster field we
obtained $68\times 420\;sec$ and $9\times 40\div90\;sec$ exposures in $V$,
$5\times 600\;sec$ and $4\times 80\div120\;sec$ exposures in $B$. For the
comparison field we collected $53\times 420\;sec$ and $3\times 60\div170\;sec$
exposures in $V$, $2\times 600\;sec$ and one $120\;sec$ exposure in $B$. 

\section{Data reduction}

The preliminary processing of the CCD frames was done with the standard
routines in the IRAF-CCDPROC package.\footnote{IRAF is distributed by
the National Optical Astronomy Observatories, which are operated by
the Association of Universities for Research in Astronomy, Inc.,
under cooperative agreement with the NSF} Due to imperfect mirror alignment 
stars on some parts of the frame had severely elongated point spread
functions (PSF). In order not to deteriorate the quality of the photometry
over the good parts of the frame the leftmost 500 columns were masked out
with the IMREPLACE routine. 

Photometry was extracted using the {\it Daophot/Allstar} package (Stetson
1987). A PSF varying quadratically with the position on the frame was used.
Due to the large pixel scale, resulting in marginal sampling of stellar
profiles, the PSF was modeled with a Gaussian function. Stars were
identified using the FIND subroutine and aperture photometry was done on
them with the PHOT subroutine. Around 150 bright isolated stars 
were initially chosen by {\it Daophot} for the construction of the PSF. 
Of those the stars with profile errors greater than twice the average were
rejected and the PSF was recomputed. This procedure was repeated until no
such stars were left on the list. The PSF was then further refined on
frames with all but the PSF stars subtracted from them. This procedure was
applied twice. The PSF obtained in the above method was then used by
{\it Allstar} in profile photometry.

The frames intended as templates for our databases were taken as the
average of three frames with the best seeing. A similar reduction procedure
was applied as for single frames. After having obtained the profile
photometry FIND was run again on the subtracted frame with a threshold of
10-15 sigma. Detections at a distance of less than one half FWHM from
stars identified previously were rejected, as were the stars for which PHOT
could not determine a magnitude. Upon closer examination the latter proved
to be spurious peaks caused by the subtraction of the imperfect PSF,
especially on the edges of the frames, where the PSF was deformed. Allstar
was run again on the expanded list, which resulted in the rejection of some
more objects, and then once more, to refine the positions of stars. 

For many of the bright stars which were saturated on the template useful
photometry could be obtained on most of the other images with inferior
seeing. The brightest stars from a short exposure were selected and added
to the template list. Great care was taken to remove false detections in
the profiles of saturated stars from the final list. 

The final template star list was then transformed to the $(X,Y)$ coordinate
system of each of the frames and used as input to {\it Allstar} in the
fixed-position mode. The output profile photometry was transformed to the
common instrumental system of the template image and then combined into a
database. The databases were created for the $420\;sec$ exposures in the
$V$ filter only in each of the two fields. 

Equatorial coordinates were determined for all objects found on the $V$
filter templates. The transformation from rectangular to equatorial
coordinates was derived using 4045 and 3402 reference stars from the
USNO-A2 catalog (Monet et al. 1996) for the cluster and the comparison
field respectively. In both cases the average difference between the
catalog and the computed coordinates was less than $0.5\;\arcsec$ in right
ascension and $0.25\;\arcsec$ in declination. 

\section{Selection of variables}

We have followed the procedure for selecting variables given in 
Kaluzny et al. (1998), where it is described in detail. Here, only a brief
summary will be presented.

The reduction procedure described in the previous section
produces databases of calibrated $V$ magnitudes and their standard
errors. The $V$ databases for the cluster and the comparison fields
contained 10705 and 8143 stars, with up to 68 and 54 measurements,
respectively. 

Measurements flagged as ``bad'' (with unusually large {\em
Daophot} errors, compared to other stars) and measurements with errors
exceeding the average error, for a given star, by more than $4\sigma$
were removed. For further analysis we used only those stars that had
at least $N_{good}>N_{max}/2\;$ measurements. There are 10153 and 7664
such stars in the $V$ databases of the cluster and comparison fields,
respectively.

Our next goal was to select a sample of variable stars from the total
sample defined above. There are many ways to proceed, and we largely
followed the approach of Stetson (1996), also described in Kaluzny et
al.(1998).  For each star we computed the Stetson's variability index $J_S$ 
and stars with values exceeding some minimum value $J_{S,min}$ were
considered candidate variables.  The definition of $J_S$ is rooted in the
assumption that on each visit to the program field at least one pair of
observations is obtained, and only when both observations have a residual
from the mean of the same sign does the pair contribute positively to the
variability index.  The definition of Stetson's variability index includes
the standard errors of individual observations.  If, for some reason, these
errors were over- or underestimated, we would either miss real variables, or
select spurious variables as real ones. Using the procedure described
in Kaluzny et al. (1998), we scaled the {\em Daophot} errors to better
represent the ``true'' photometric errors.  We then selected the candidate
variable stars by computing the value of $J_S$ for the stars in our $V$
database.  We used a cutoff of $J_{S,min}=0.75$ to select 277 and 284
candidate variable stars in the cluster and the comparison fields,
respectively. 

The light curves of all variable candidates were examined, as were the
images of the stars themselves on the CCD frames, to weed out false
detections in the wings of bright stars, bad pixels, etc. After the
rejection of spurious variables and stars with noisy/chaotic light curves
we were left with 31 variables in the cluster field and 14 in the
comparison field.  

The periodicities for all candidate variables were analyzed using a
variant of the Lafler-Kinman (1965) string-length technique proposed by
Stetson (1996). The periods were then refined using the analysis of
variance method, as described by Schwarzenberg-Czerny (1989).

\section{Color-magnitude diagrams}

To construct the color-magnitude diagrams we used the $V$ photometry 
from the templates. To extend the CMD to lower magnitudes we added bright
stars from our best quality short exposures. The $B$ filter photometry for
the cluster field was obtained from an image constructed from three long
exposures with the best seeing, just as it was done for the $V$ templates. 
Bright stars selected from a short $B$ exposure were appended to the 
photometry list. In case of the comparison field only two long exposures in
the $B$ filter were available, taken far apart in time and under different
seeing conditions. The photometry lists from the two images were merged,
with the better of the two measurements retained for each star present on
both lists. By better we mean the magnitude determination with a smaller
standard error, as computed by {\it Allstar}.

For the variable stars we have determined their $V$ magnitudes and $B-V$
colors following a different procedure. In case of the eclipsing binaries we
used the magnitude outside of the eclipses, $V_{max}$ to position them on
the CMD. For the pulsating stars we determined their average magnitude
$\langle V \rangle$. The variables for which we did not derive a period
were plotted with the maximum magnitude $V_{max}$ they attained during our
observing run. 

To determine the colors of the variables the $B$ magnitudes from the 600 
sec. exposures were used. The $V$ magnitudes were interpolated from two
$V$ exposures, one taken before and one after, to the beginning of the B
exposure. The average $B-V$ colors were computed from four color
determinations in the cluster field and two in the comparison field. The
scatter of the color determinations for most variables brighter than 17th
mag. was about 0.02 mag. Some of the eclipsing variables displayed a wider
range of color variations, up to 0.2 mag., but that seems to be an
intrinsic property of those systems, most likely caused by the different
colors of the components of the detached systems or gravitational darkening
in the semi-detached and contact pairs. 

The transformation of the instrumental magnitudes to the standard system
was derived from observations of the Landolt fields (Landolt 1992). The
following relations were adopted: 
\begin{eqnarray*}
v   = V - 0.002\times(B-V) + const\\
b-v =     0.966\times(B-V) + const
\end{eqnarray*}
The zero points were determined from the unpublished photometry obtained by
the second author on the KPNO 2.1m telescope in 1992.

\begin{figure}[tbp]
\plotfiddle{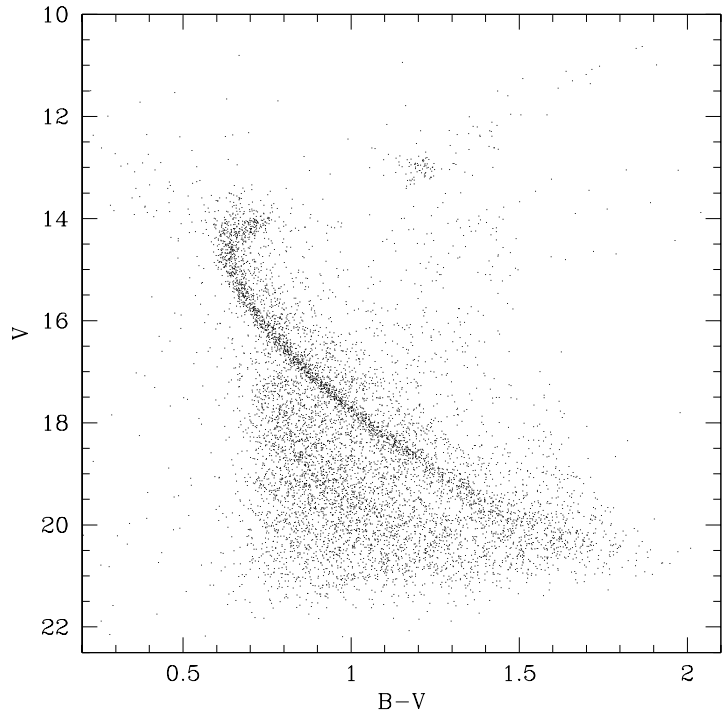}{8cm}{0}{50}{50}{-160}{-85}
\caption{The color-magnitude diagram for the field centered on the NGC 7789
open cluster.}
\label{fig:cmd}
\end{figure}

\begin{figure}[tbp]
\plotfiddle{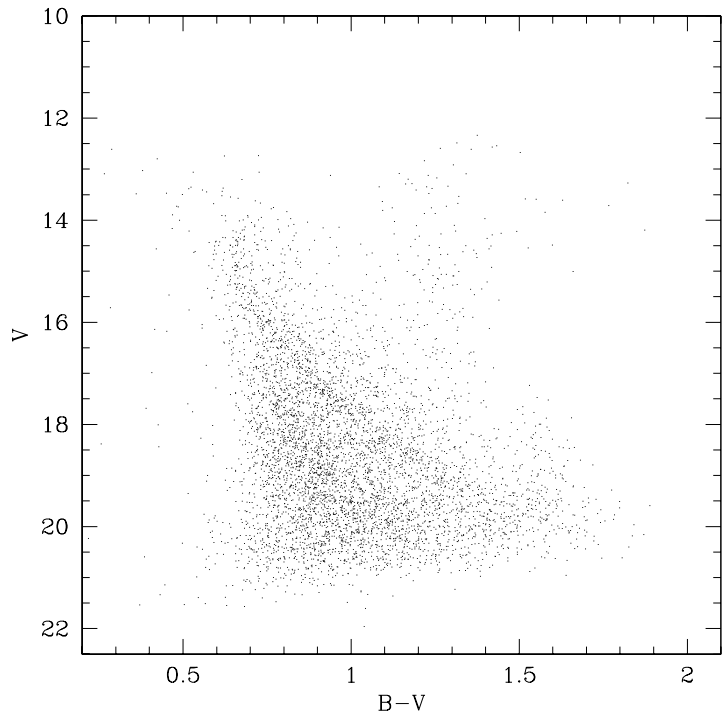}{8cm}{0}{50}{50}{-160}{-85}
\caption{The color-magnitude diagram for the comparison field.}
\label{fig:cmdc}
\end{figure}

\begin{figure}[tbp]
\plotfiddle{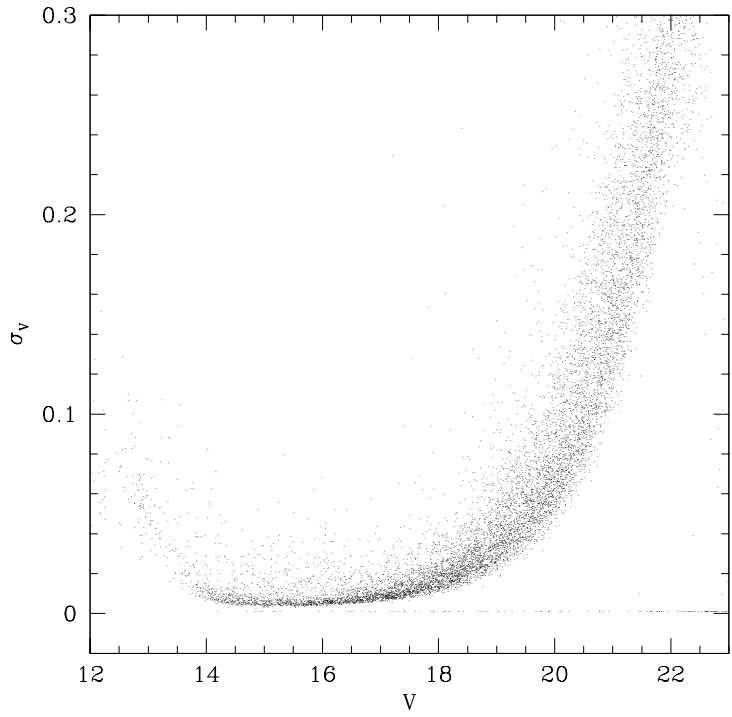}{8cm}{0}{50}{50}{-160}{-85}
\caption{The average standard $V$ magnitude errors plotted as a function of
$V$ for stars in the cluster field.}
\label{fig:v_s}
\end{figure}

\begin{figure}[tbp]
\plotfiddle{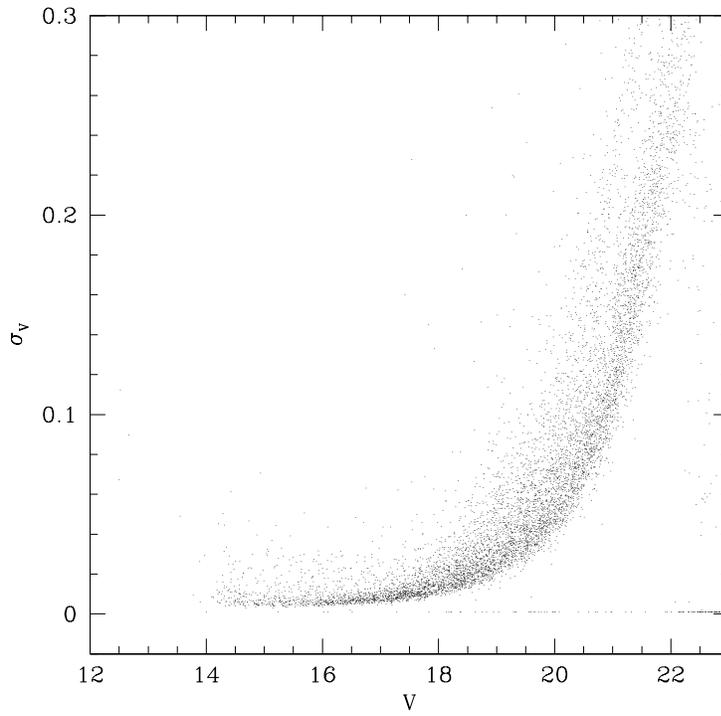}{8cm}{0}{50}{50}{-160}{-85}
\caption{The average standard $V$ magnitude errors plotted as a function of
$V$ for stars in the comparison field.}
\label{fig:vc_s}
\end{figure}

The color-magnitude diagrams for the field centered on the cluster and the
comparison field are shown in Figures~\ref{fig:cmd} and \ref{fig:cmdc},
respectively. The diagram for the cluster field (Fig.~\ref{fig:cmd}) shows
a well defined main sequence that can be traced down to $V\sim20.5$ mag,
and a substantial number of blue straggler candidates at its extention to
lower magnitudes, as well as an extended red giant branch and a
reasonably well-defined red clump. There is no obvious indication of a post
main sequence gap. The comparison field CMD (Fig. ~\ref{fig:cmdc}) 
consists mainly of field stars, although the cluster main sequence is still 
discernible. We were able to trace it out easily to a distance of
$\sim15\arcmin$ from the cluster center. The depth of the CMDs is biased by
the $B$ band photometry. Figures \ref{fig:v_s} and \ref{fig:vc_s} show the
average standard $V$ magnitude errors plotted as a function of $V$.

\begin{figure}[tbp]
\plotfiddle{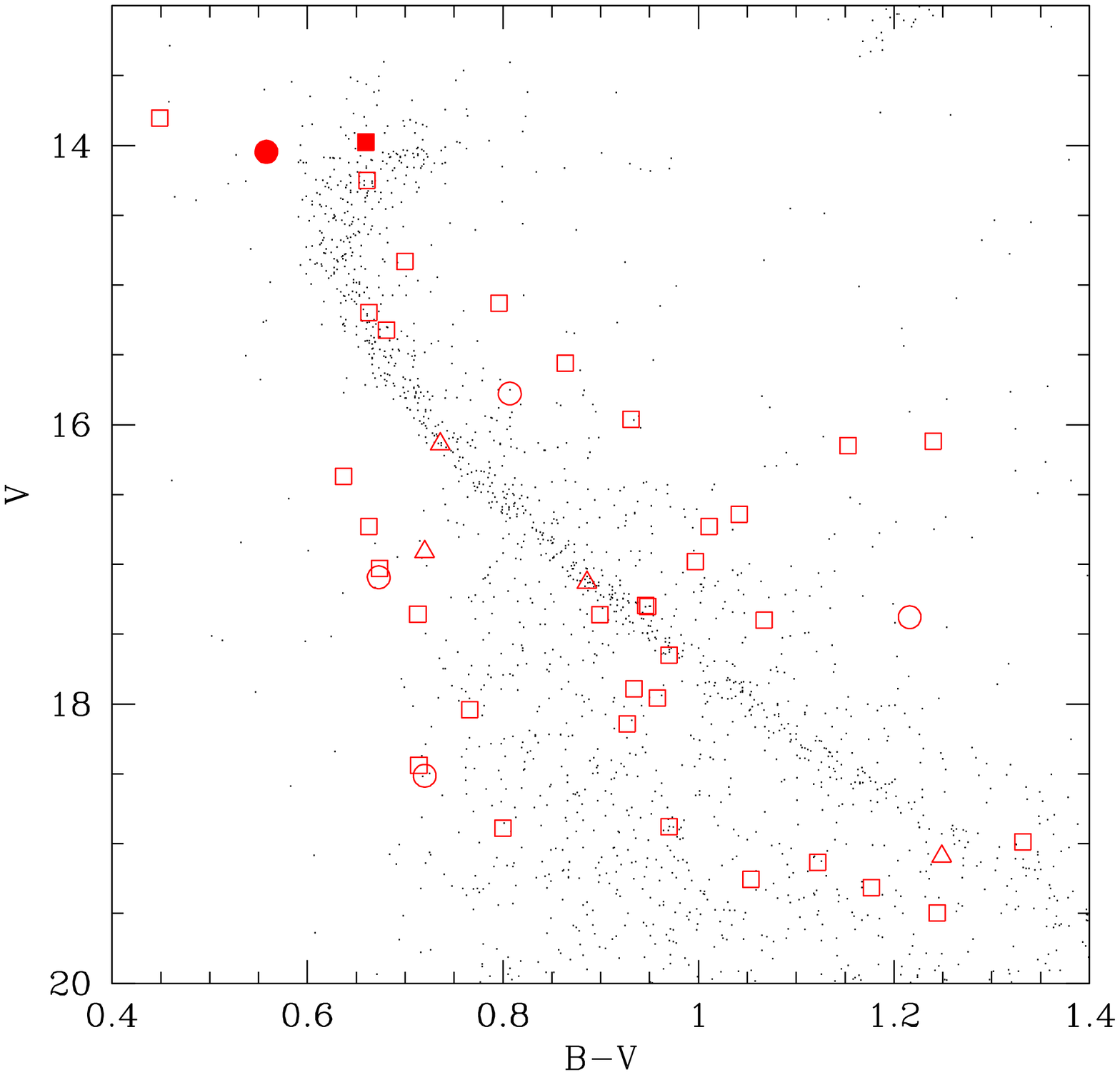}{8.5cm}{0}{50}{50}{-160}{-85}
\caption{The color-magnitude diagram for the variable stars. Eclipsing
binaries are marked with squares, pulsating variables with circles and
miscellaneous variables with triangles. Solid symbols indicate confirmed
cluster members. Stars within a radius of 500 pixels from the cluster
center are also plotted.} 
\label{fig:cmdvar}
\end{figure}

The color-magnitude diagram for the variable stars is presented in
Figure~\ref{fig:cmdvar}. Each type of variables is denoted by a different
symbol: eclipsing binares by squares, pulsating variables by circles and
miscellaneous by triangles. Stars within a radius of 500 pixels
($\sim6\arcmin$) from the cluster center are plotted in the
background. Solid symbols indicate cluster members confirmed by proper
motion measurements (McNamara \& Solomon 1981). The brightest and bluest
variable, V9097, is a known non-member with $0\%$ probability of
belonging to the cluster. Variables located close (at less than about
$6\arcmin$) to the center of the cluster show a tendency to group on and
above the main sequence. 

\section{Catalog of variables}

In this section we present the light curves of the 45 variables found
in the cluster and the comparison fields. The following convention was
adopted for the naming of variable stars: letter V for ``variable'' and the
number of the star in the $V$ filter database. For variables found in the
comparison field a C is appended to the variable name. Tables
\ref{tab:ecl}, \ref{tab:pul}, \ref{tab:oth} list the variables sorted by
three categories: eclipsing binaries, pulsating variables and other
variables for which we could not obtain a period determination. The light
curves of the variables are shown in Figures~\ref{fig:ecl}, \ref{fig:pul}
and \ref{fig:oth}. The $60\arcsec\times60\arcsec$ finding charts are
provided in Figure \ref{fig:map}.

\subsection{Eclipsing binaries}

We have identified 35 eclipsing binaries in both fields, 26 in the cluster
field and nine in the comparison field, 26 of them newly discovered. 
In addition, two of the eclipsing binaries from the cluster field were also
identified in the comparison field. In Table \ref{tab:ecl} we present the
parameters for the eclipsing binaries, sorted by period: name, J2000.0
coordinates, period, $V_{max}$ magnitude of the system outside of the
eclipse and the average $B-V$ color. Their phased light curves are shown in
Figure~\ref{fig:ecl}. 

Most of the eclipsing binaries have periods shorter than one day and
exhibit W UMa type light curves. 

The light curves of two variables, V4276 and V6698, show
variations similar to those of RS Canum Venaticorum stars. Another
variable, V2130, a contact binary, also displays strong assymetry of the
light outside of the eclipses: the system achieves maximum brightness
between the primary and the secondary eclipse. More observations would be
needed to see if and how the light curves evolve with time, to gain insight
into the true nature of these systems. 

The cataclysmic variable candidate V9, found by Jahn et al. (1995),
identified by us as V3283, displays brightness variations outside of the
eclipses with an amplitude of 0.03-0.05 mag. on timescales of 1-2 hours,
very much like those described in that paper. 

Two of our eclipsing binaries have their membership probabilities
determined from proper motion analysis by McNamara \& Solomon. 
One of them, V3785 is a cluster member with probability $P=0.98$ while
the other, V9097, is not ($P=0.0$).

We extracted from our database the light curve for variable V6 found by
Jahn et al. (1995). We noticed traces of sinusoidal variability when phased
with the published period, but our light curve was too noisy to obtain an
independent period determination. We therefore decided not to include this
variable into our catalog. 

It should be noted that our classification should not be regarded as
final. In some cases it was difficult to determine whether a variable is an
eclipsing binary or a pulsating variable with half the period. One example
is V820C, which exhibits maxima that may seem too sharp for an eclipsing 
binary, but was still classified as such because of a slight asymmetry of
the minima and and its location on the CMD. Information as to the cluster
membership would be helpful in many cases. More observations would be
needed for some objects, to obtain a better light curve and refine the
color determination. 

\begin{figure}[t]
\plotfiddle{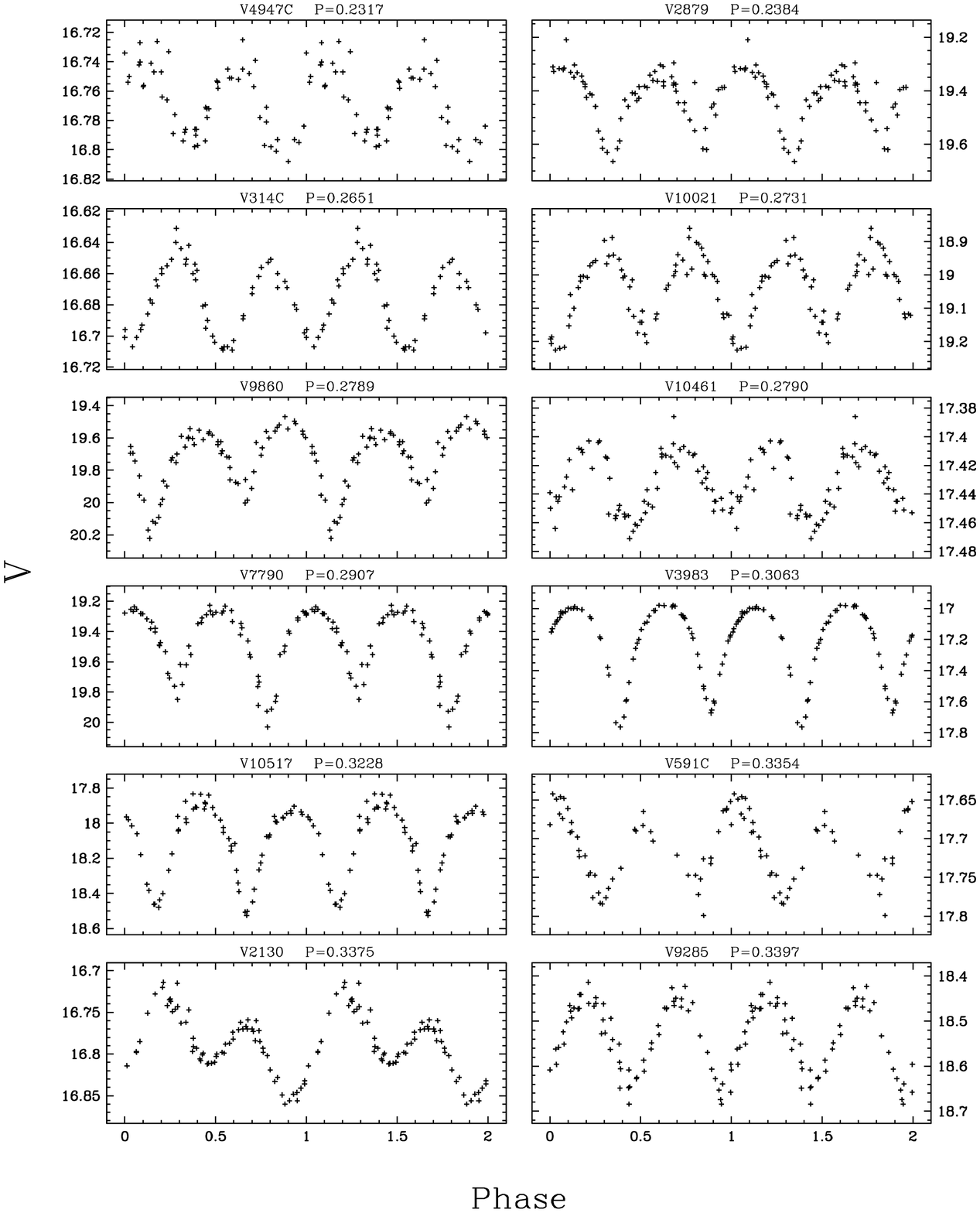}{19.5cm}{0}{83}{83}{-260}{-40}
\caption{Phased $V$ filter light curves of the eclipsing binaries.}
\label{fig:ecl}
\end{figure}

\addtocounter{figure}{-1}
\begin{figure}[p]
\plotfiddle{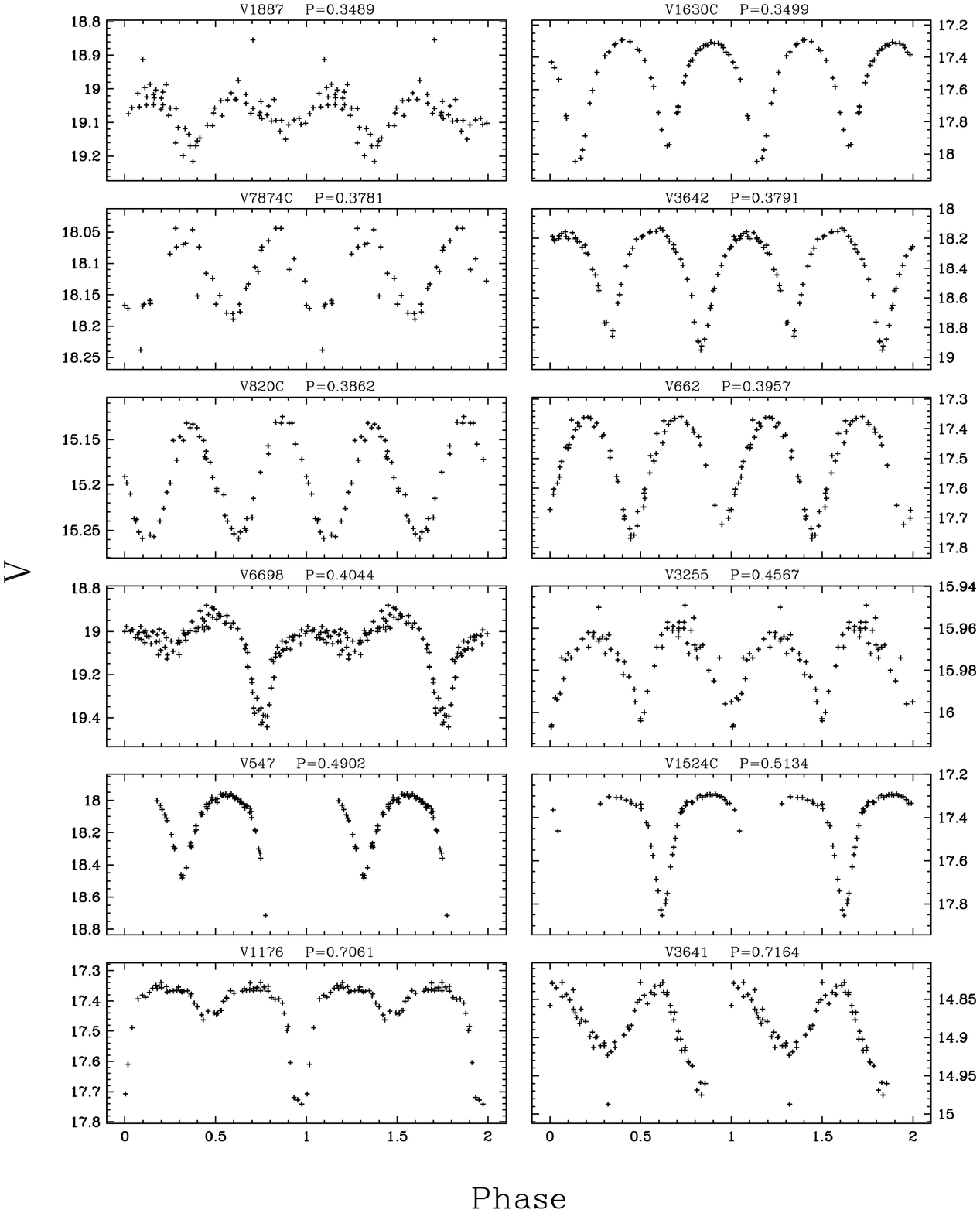}{19.5cm}{0}{83}{83}{-260}{-40}
\caption{Continued.}
\end{figure}

\addtocounter{figure}{-1}
\begin{figure}[p]
\plotfiddle{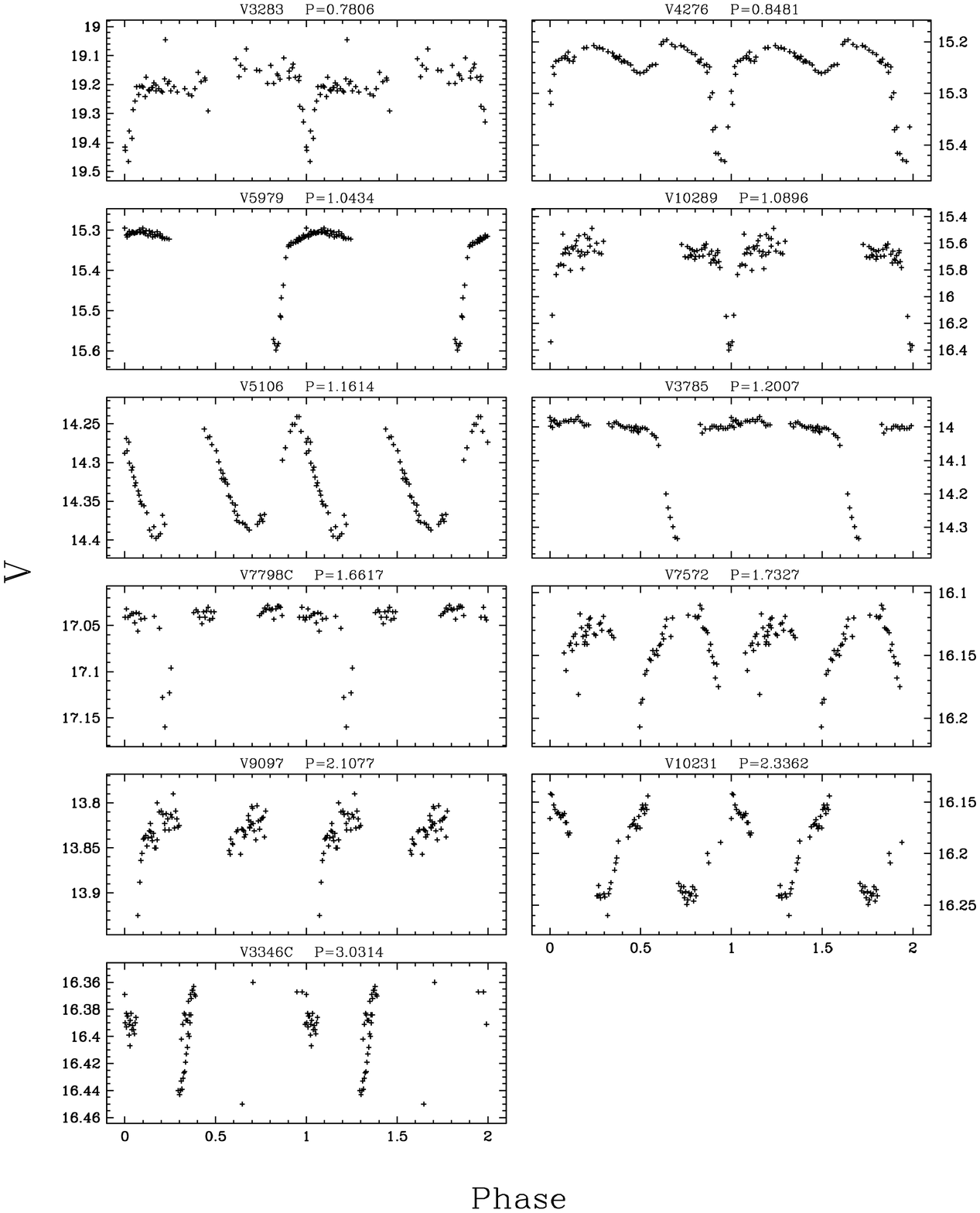}{19.5cm}{0}{83}{83}{-260}{-40}
\caption{Continued.}
\end{figure}

\begin{table}
\caption[]{\sc Eclipsing Binaries in NGC 7789\\}
\begin{tabular}{lllllllll}
\hline\hline
Name & $RA(2000)$ & $Dec(2000)$ & $P$ (days) & $V_{max}$ & 
$\langle B-V \rangle$ & \multicolumn{2}{l}{Comments} \\
\hline
V4947C  & 23 59 23.01 & 56 35 51.05 & 0.2317 & 16.73 & 0.66 & EW & \\
V2879   & 23 57 57.25 & 56 45 48.38 & 0.2384 & 19.32 & 1.18 & EW & V5 \\
V314C   & 23 58 46.38 & 56 46  2.47 & 0.2651 & 16.64 & 1.04 & EW & \\
V10021  & 23 56 18.29 & 56 34 14.88 & 0.2731 & 18.89 & 0.80 & EW & \\
V9860   & 23 57 25.21 & 56 34 36.91 & 0.2789 & 19.49 & 1.24 & EB & \\
V10461  & 23 57 10.65 & 56 33 27.07 & 0.2790 & 17.40 & 1.07 & EW & \\
V7790   & 23 58 12.06 & 56 38  6.15 & 0.2907 & 19.25 & 1.05 & EW & \\
V3983   & 23 56 38.70 & 56 43 58.76 & 0.3063 & 16.98 & 1.00 & EW & \\
V10517  & 23 57 34.70 & 56 33 20.30 & 0.3228 & 17.89 & 0.93 & EW & \\
V591C   &  0  0 39.42 & 56 45 29.64 & 0.3354 & 17.65 & 0.97 & EW & \\
V2130   & 23 57 49.41 & 56 46 59.91 & 0.3375 & 16.73 & 1.01 & EW & V4 \\
V9285   & 23 56 15.36 & 56 35 34.72 & 0.3397 & 18.44 & 0.71 & EW & \\
V1887   & 23 58 13.47 & 56 47 25.70 & 0.3489 & 18.99 & 1.33 & EB & \\
V1630C  & 23 59 40.91 & 56 43  7.76 & 0.3499 & 17.30 & 0.95 & EW & \\
V7874C  & 23 58 10.64 & 56 29 33.09 & 0.3781 & 18.04 & 0.77 & EW & \\
V3642   & 23 56 35.86 & 56 44 30.00 & 0.3791 & 18.14 & 0.93 & EW & \\
V820C   & 23 59 50.81 & 56 44 55.72 & 0.3862 & 15.13 & 0.80 & EW & \\
V662    & 23 56 50.03 & 56 49 30.42 & 0.3957 & 17.36 & 0.90 & EW & \\
V6698   & 23 58 37.68 & 56 39 53.80 & 0.4044 & 18.88 & 0.97 & EB & \\
V3255   & 23 57 24.45 & 56 45 13.00 & 0.4567 & 15.96 & 0.93 & EW & V7 \\
V547    & 23 56 44.55 & 56 49 44.23 & 0.4902 & 17.96 & 0.96 & EB & \\
V1524C  & 23 59 33.48 & 56 43 23.88 & 0.5134 & 17.29 & 0.95 & EW?& \\
V1176   & 23 56 56.38 & 56 48 34.79 & 0.7061 & 17.36 & 0.71 & EA & \\
V3641   & 23 57 33.45 & 56 44 32.95 & 0.7164 & 14.83 & 0.70 & EB & V2 \\
V3283   & 23 57  0.19 & 56 45  8.47 & 0.7806 & 19.13 & 1.12 & EA & V9 \\
V4276   & 23 57 25.68 & 56 43 33.22 & 0.8481 & 15.20 & 0.66 & EB & V8 \\
V5979   & 23 57 39.45 & 56 41  0.07 & 1.0434 & 15.32 & 0.68 & EA & V3 \\ 
V10289  & 23 56  9.06 & 56 33 42.92 & 1.0896 & 15.56 & 0.86 & EA & \\
V5106   & 23 57  9.92 & 56 42 18.08 & 1.1614 & 14.25 & 0.66 & EW & V1 \\
V3785   & 23 56 32.15 & 56 44 17.60 & 1.2007 & 13.98 & 0.66 & EA & \\
V7798C  & 23 59 21.52 & 56 29 48.39 & 1.6617 & 17.03 & 0.67 & EA & \\
V7572   & 23 56 50.81 & 56 38 26.21 & 1.7327 & 16.12 & 1.24 & EA & V12 \\
V9097   & 23 57 38.60 & 56 35 58.04 & 2.1077 & 13.80 & 0.45 & EA & \\
V10231  & 23 57 44.85 & 56 33 55.63 & 2.3362 & 16.15 & 1.15 & EW & \\
V3346C  & 23 59  3.73 & 56 39 17.32 & 3.0314 & 16.37 & 0.64 & EA & \\
\hline
\end{tabular}
\label{tab:ecl}
\end{table}

\subsection{Pulsating variables}

\begin{figure}[t]
\plotfiddle{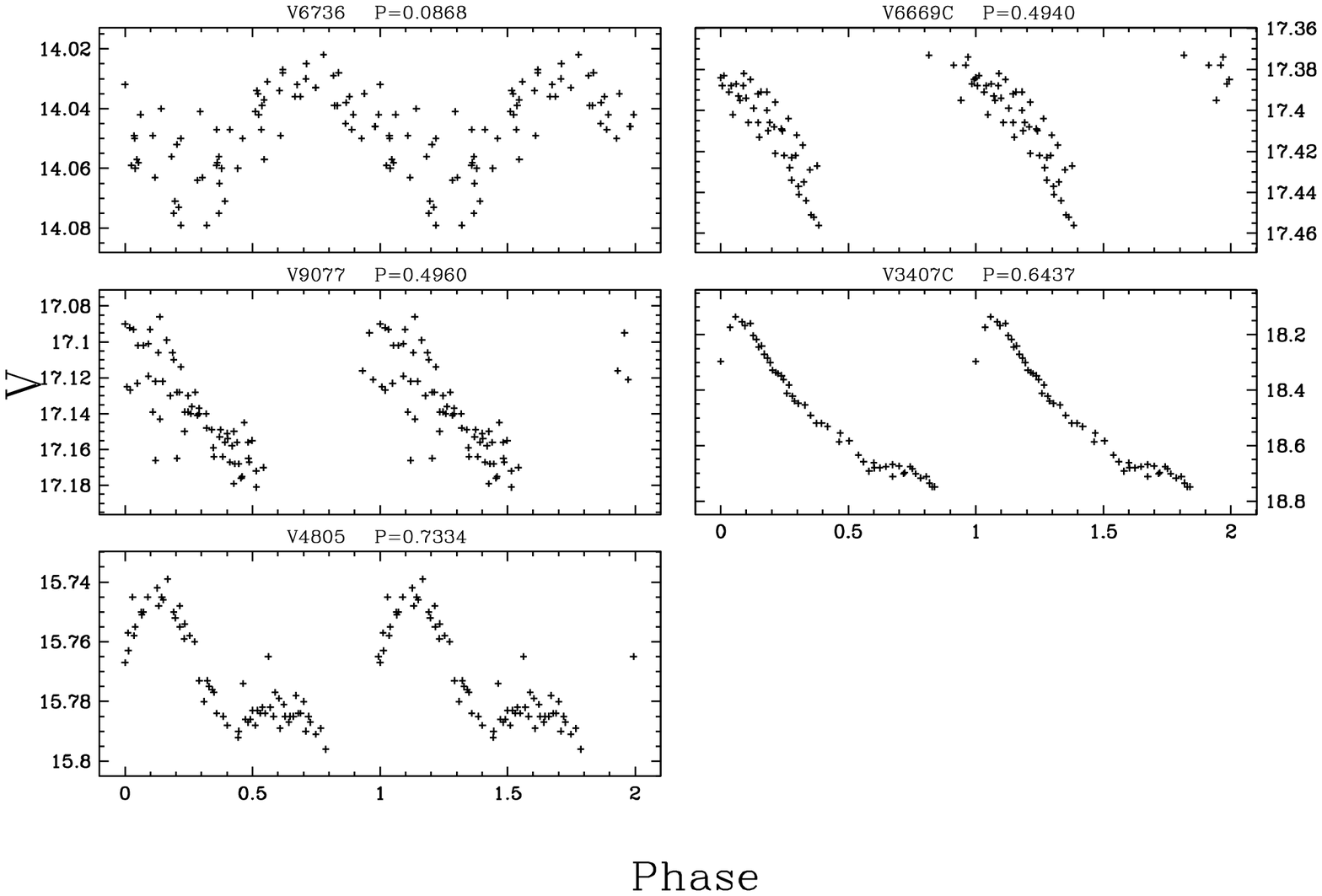}{1.5cm}{0}{83}{83}{-245}{-300}
\caption{Phased $V$ filter light curves of the pulsating variables.}
\label{fig:pul}
\end{figure}

We have found five pulsating variables, three in the cluster field and two
in the comparison field, one of them already known. In Table
\ref{tab:pul} we  present the parameters for the pulsating variables,
sorted by period: name, J2000.0 coordinates, period, the mean $V$ magnitude
and the $\langle B-V \rangle$ color. Their phased light curves are shown in
Figure~\ref{fig:pul}.

The variable V3407C is a background RRab Lyrae star. V4805 may
also be an RR Lyrae variable. V6736 is most likely a $\delta$ Scuti
variable and a cluster member with $P=0.95$ (McNamara \& Solomon 1981). 

For variables V6669C and V9077 on each night we observed a fall in
brightness of about 0.1 mag. We phased them with a period close to 
0.5 of a day, but due to an incomplete phase coverage this is not a very
reliable estimate.

\begin{table}
\caption[]{\sc Pulsating Variables in NGC 7789\\}
\begin{tabular}{llllllll}
\hline\hline
Name& $RA(2000)$ & $Dec(2000)$ & $P$ (days) &  $\langle V \rangle$ & 
$\langle B-V \rangle$ & Comments & \\ 
\hline
V6736   & 23 57 30.93 & 56 39 49.31 & 0.0868 & 14.05 & 0.56&$\delta$ Sct&V10\\
V6669C  & 23 58 22.86 & 56 32  9.90 & 0.4940 & 17.38 & 1.22 & &\\
V9077   & 23 57 58.79 & 56 36  0.30 & 0.4960 & 17.09 & 0.67 & &\\
V3407C  &  0  0  6.64 & 56 39 12.11 & 0.6437 & 18.51 & 0.72 & RR Lyr&\\
V4805   & 23 57 53.72 & 56 42 44.52 & 0.7334 & 15.78 & 0.81 & RR Lyr&\\

\hline
\end{tabular}
\label{tab:pul}
\end{table}

\subsection{Miscellaneous variables}

\begin{figure}[t]
\plotfiddle{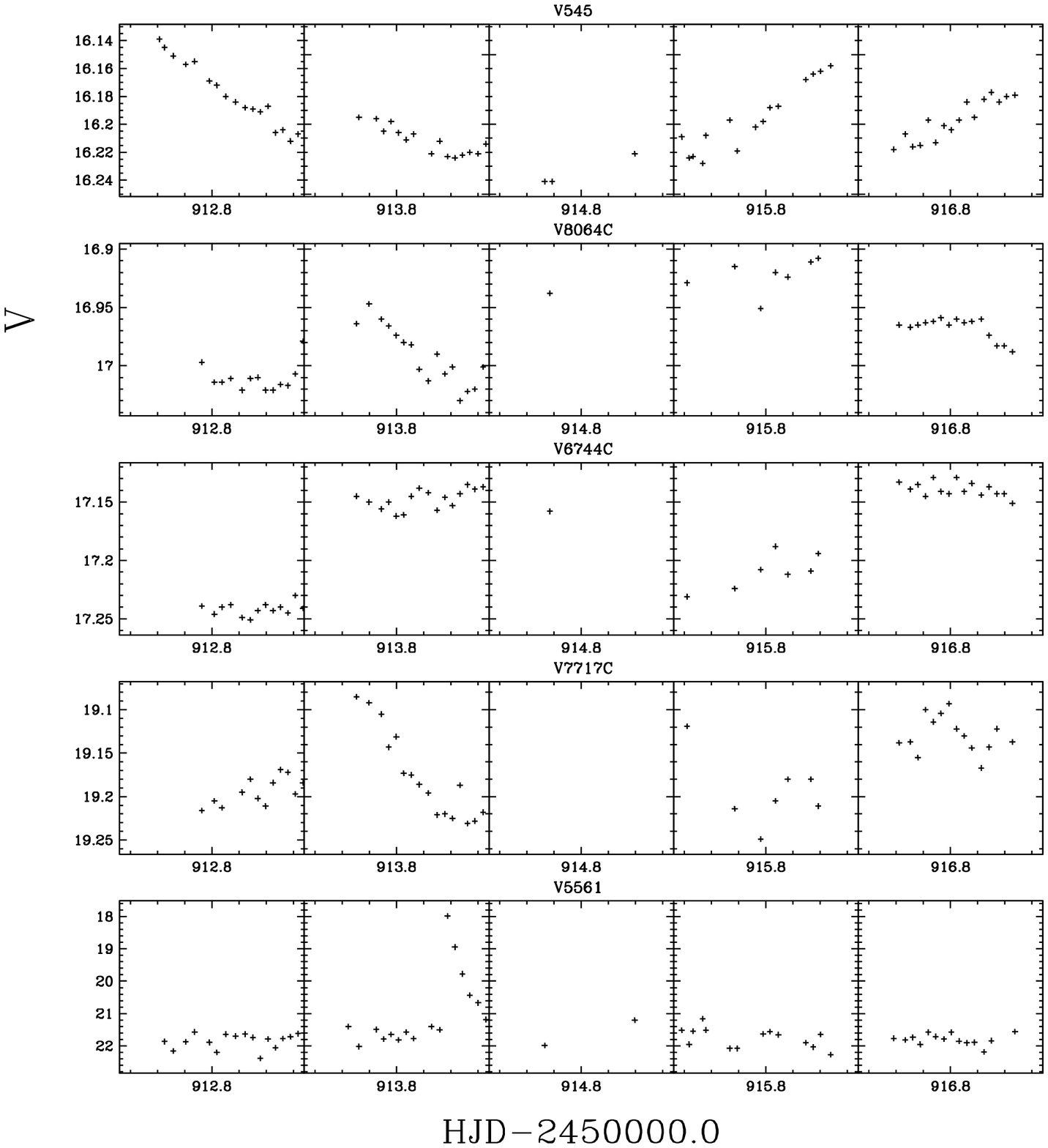}{19.5cm}{0}{83}{83}{-260}{-40}
\caption{The $V$ filter light curves of the miscellaneous variables.}
\label{fig:oth}
\end{figure}

We have identified five other variables, two in the cluster field and three
in the comparison field, all of them newly discovered. 
In table \ref{tab:oth} we present the parameters of those variables, sorted
by $V_{max}$ magnitude: name, J2000.0 coordinates, the $V_{max}$ magnitude
and the $\langle B-V \rangle$ color. Their light curves are shown in
Figure~\ref{fig:oth}. 

One variable, V5561 is probably a UV Ceti or a BY Dra variable. During five
nights of monitoring it underwent an outburst with an amplitude of 4
magnitudes on the second night of observations. 

Some of the variables listed here may be periodic, with our observations
covering only a part of the period. Due to the short timespan of our
observations and few data points obtained on the third night we were able
to determine periods shorter than about 2.5 days. 

We have detected the variability of the four previously known variables,
V11, V13, V14 and V15, found by Jahn et al. (1995). We decided not to
include them into our catalog due to the poor quality of the light curves
derived from our observations and small amplitudes of variability. 

\begin{table}
\caption[]{\sc Miscellaneous Variables in NGC 7789\\}
\begin{tabular}{llllll}
\hline\hline
Name& $RA(2000)$ & $Dec(2000)$ & $V_{max}$ & $\langle B-V \rangle $ 
& Comments\\ 
\hline
V545    & 23 57 29.98 & 56 49 46.57 & 16.14 & 0.74 & \\
V8064C  & 23 59 25.18 & 56 29 10.92 & 16.91 & 0.72 & \\
V6744C  & 23 58 18.03 & 56 31 58.74 & 17.13 & 0.89 & \\
V7717C  & 23 58 25.12 & 56 29 55.66 & 19.09 & 1.25 & \\
V5561   & 23 56  8.30 & 56 41 34.02 & 19.83&\nodata& UV Ceti \\
\hline
\end{tabular}
\label{tab:oth}
\end{table}

\section{Cluster membership of the contact binaries}

\begin{figure}[t]
\plotfiddle{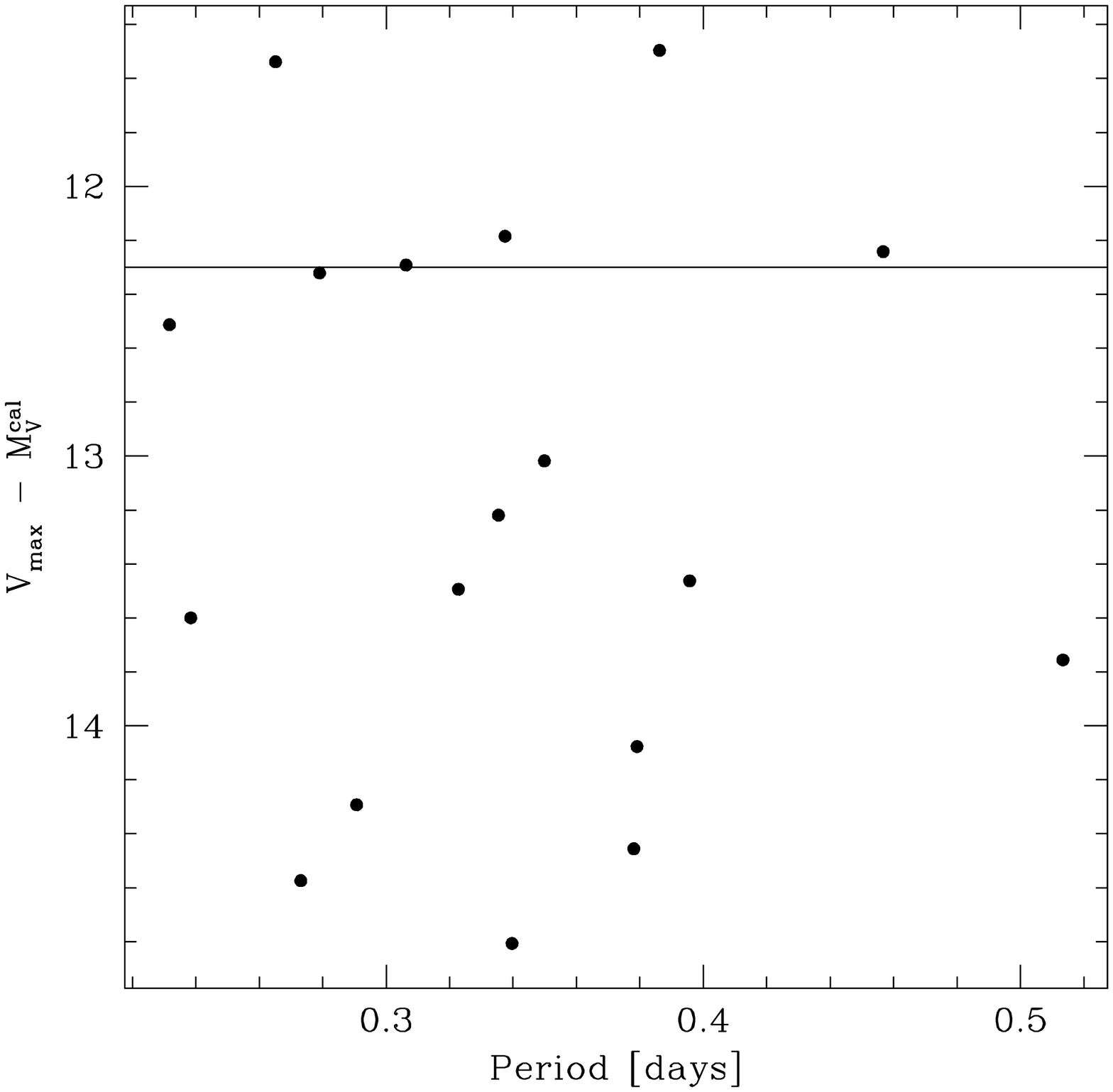}{8cm}{0}{50}{50}{-160}{-85}
\caption{The period versus apparent distance modulus diagram for the 
contact binaries found in both the cluster and the comparison field. 
The variable V5106 with an apparent distance modulus of 13.147 mag. 
is not shown in the plot, due to an outlying period.}
\label{fig:mv}
\end{figure}

In an attempt to asess the cluster membership of the contact binaries we 
have applied the absolute magnitude calibration established by Rucinski
\&  Duerbeck (1997) to calculate their $M_V$. The adopted calibration 
expresses $M_V$ as a function of the period and the unreddened color 
$(B-V)_0$:
\begin{equation}
M_V = -4.44 \cdot log P + 3.02 \cdot (B-V)_0 + 0.12
\end{equation}
We assumed a uniform value of reddening $E(B-V) = 0.24$ mag. across the face
of the cluster. The weighted mean deviation for the stars used to derive
the calibration was $\sigma=0.22$ mag. The calibration has not been applied 
to V10231, tentatively classified as a contact binary, due to a period 
outside of the validity range for the calibration. The apparent distance 
modulus was calculated for each system as the difference between its 
$V_{max}$ and $M_{V}^{cal}$ magnitudes. Figure \ref{fig:mv} shows the 
period versus apparent distance modulus diagram for the contact binaries 
found in both the cluster and the comparison field. The variable V5106 with 
an apparent distance modulus of 13.147 mag. is not shown in the plot, due 
to an outlying period. 

Five systems, V10461, V3983, V2130, V3255 and V4947C seem to be cluster 
members, with computed distance moduli within about $\pm0.2$ of the 
cluster's distance modulus. There seem to be no other variables within 
0.7 mag. in front of or behind the cluster. Only two variables, V314C 
and V820C are foreground systems, while the rest of the W-UMa type 
systems are located behind the cluster.

\section{Conclusions}

Our search for variable stars in the open cluster NGC 7789 has resulted in
the discovery of 45 variables: 35 eclipsing binaries, five pulsating
variables and five miscellaneous variables for which we could not determine a
period from our short span observations. Among those we have found four
interesting candidates for in-depth photometric and spectroscopic studies: 
two RS CVn eclipsing binaries, V4276 and V6698, one possible cataclysmic
variable, V3283, and one system of an unclear nature, V2130. 

Only about twice as many variables were found in the field centered on the 
cluster as in the comparison field (31 as opposed to 14). This implies that
objects not associated physically with the cluster may have a significant 
share in the total number of variables found in the cluster field, as well
as in other globular and open clusters observed on a dense
background/foreground of disk stars. 

\acknowledgments{We would like to thank Krzysztof Z. Stanek for his 
error scaling, database manipulation and period finding programs and
Grzegorz Pojma{\'n}ski for $lc$ - the light curve analysis utility,
incorporating the analysis of variance algorithm. BJM was supported
by the polish KBN grant 2P03D01416 to Grzegorz Pojma{\'n}ski. JK was 
supported by the polish KBN grant 2P03D00317 and by the NSF grant 
AST-9528096 to Bohdan Paczynski.}

\begin{figure}[t]
\plotfiddle{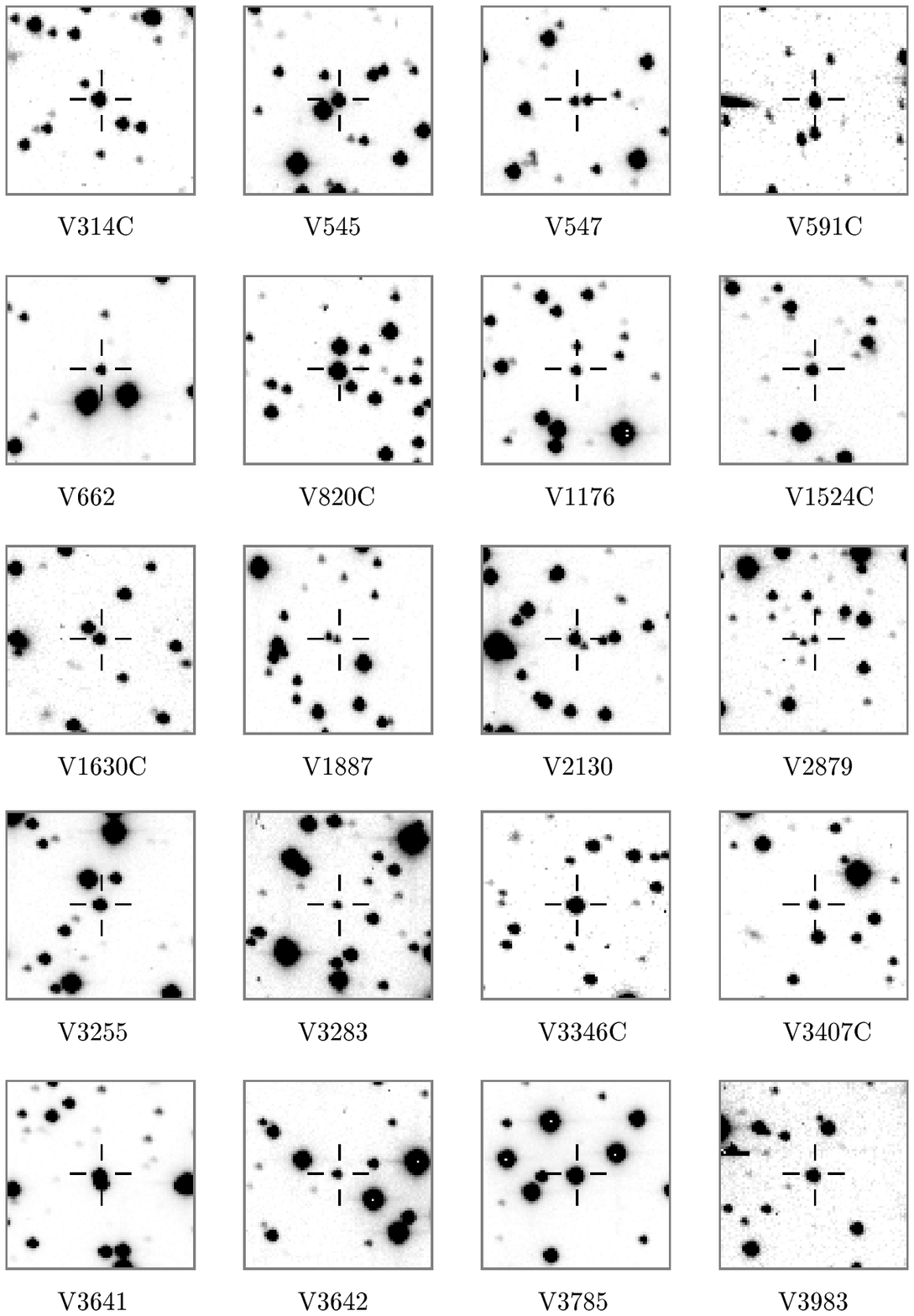}{19.5cm}{0}{100}{100}{-260}{-145}
\caption{Finding charts for the cluster and comparison field variables,
sorted by ID. North is up and east is to the left.}
\label{fig:map}
\end{figure}

\addtocounter{figure}{-1}
\begin{figure}[p]
\plotfiddle{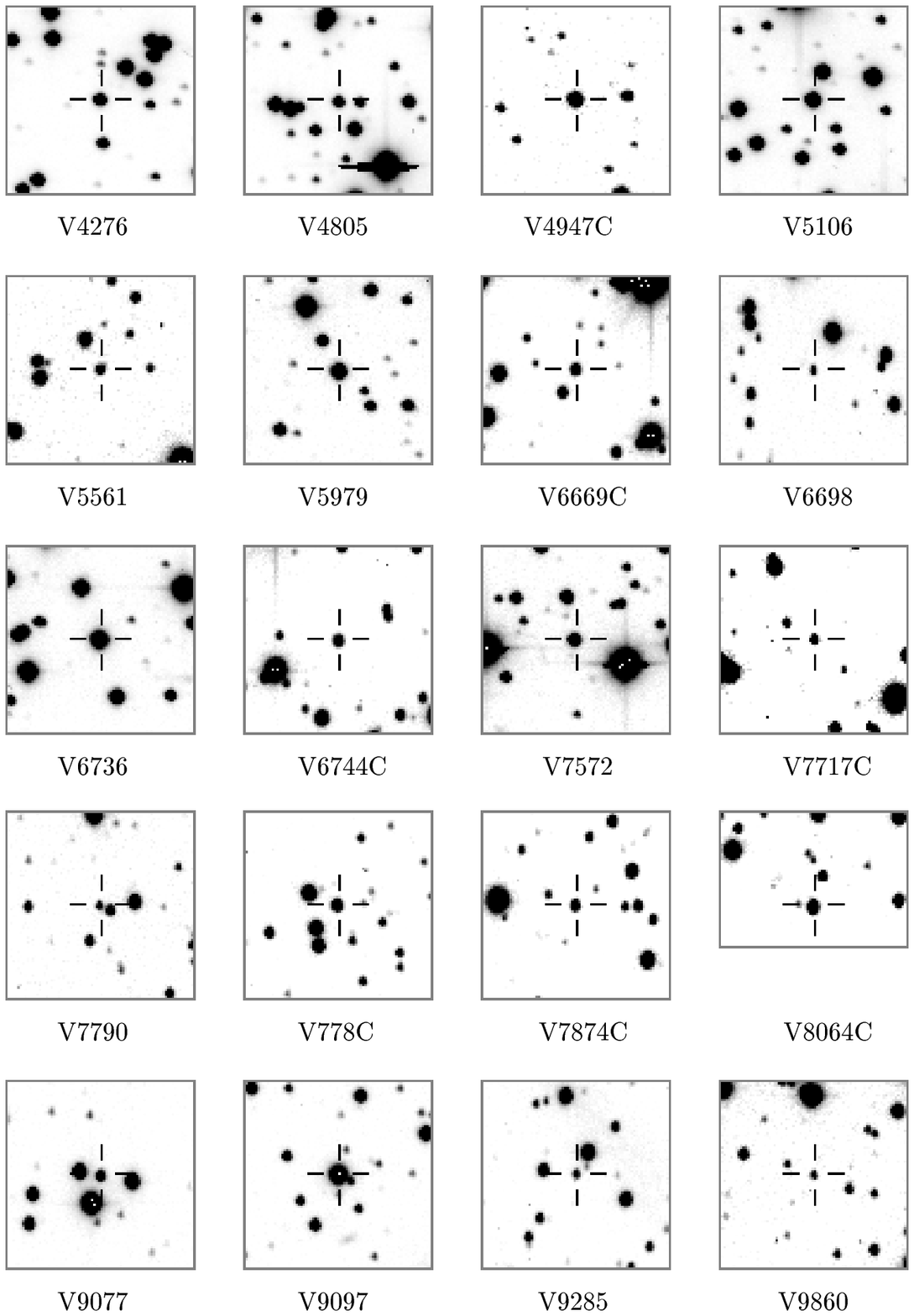}{19.5cm}{0}{100}{100}{-260}{-140}
\caption{Continued.}
\end{figure}

\addtocounter{figure}{-1}
\begin{figure}[p]
\plotfiddle{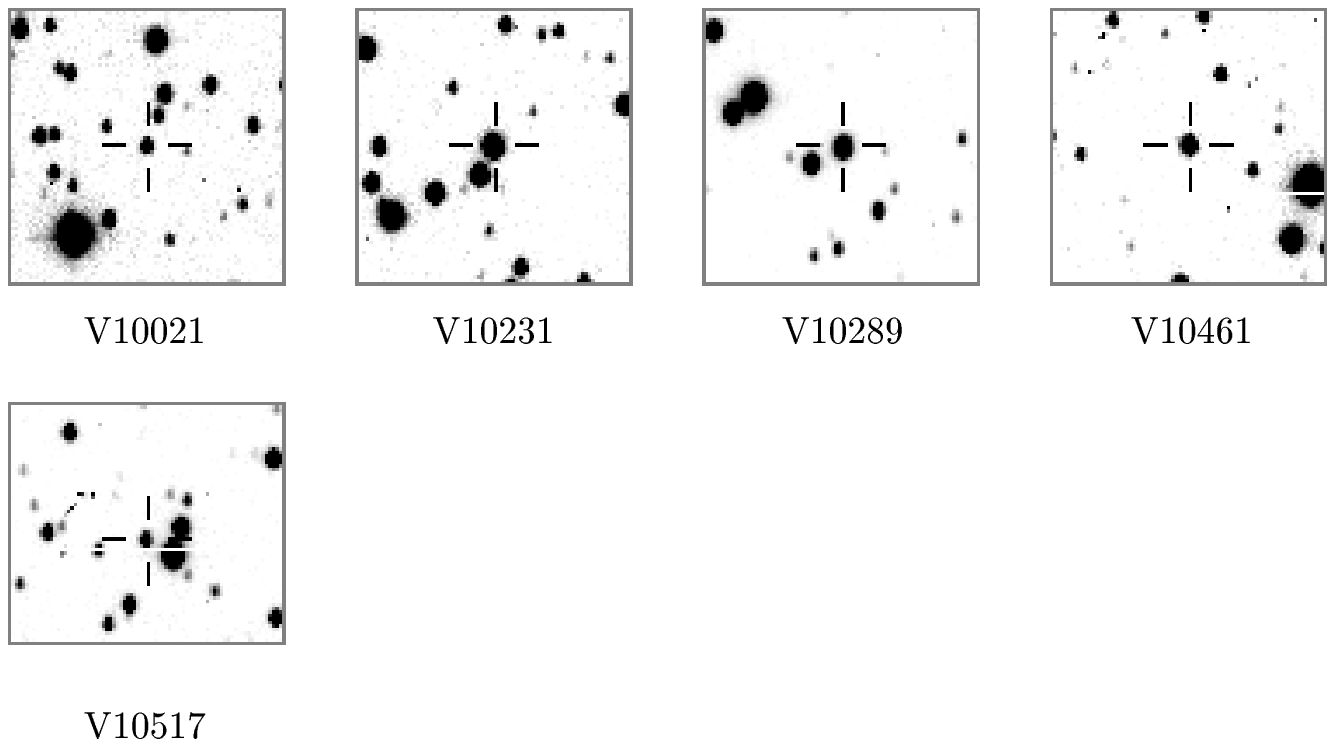}{3cm}{0}{100}{100}{-260}{-380}
\caption{Continued.}
\end{figure}
 
\end{document}